\documentstyle[12pt]{article}
\newcommand{\be}{\begin{equation}}
\newcommand{\ee}{\end{equation}}
\newcommand{\bear}{\begin{eqnarray}}
\newcommand{\ear}{\end{eqnarray}}

\newcommand{\slp}{\raise.15ex\hbox{$/$}\kern-.57em\hbox{$p$}}
\newcommand{\slv}{\raise.15ex\hbox{$/$}\kern-.57em\hbox{$v$}}
\newcommand{\slet}{\raise.15ex\hbox{$/$}\kern-.57em\hbox{$\eta$}}
\newcommand{\slB}{\raise.15ex\hbox{$/$}\kern-.57em\hbox{$B$}}
\newcommand{\slb}{\raise.15ex\hbox{$/$}\kern-.57em\hbox{$b$}}
\newcommand{\slW}{\raise.15ex\hbox{$/$}\kern-.57em\hbox{$W$}}

\begin{document}
\begin{titlepage}
\begin{flushright}
HD-THEP-99-35
\end{flushright}
\quad\\
\vspace{1.8cm}

\begin{center}
{\Large Are the neutrino masses and mixings closely related to the masses
and mixings of quarks?\footnote{Talk given
at the Johns Hopkins Workshop, Baltimore, June 10-12, 1999}}\\
\vspace{2cm}
Berthold Stech\footnote{e-mail: B.Stech@thphys.uni-heidelberg.de}\\
\bigskip
Institut  f\"ur Theoretische Physik\\
Universit\"at Heidelberg\\
Philosophenweg 16, D-69120 Heidelberg\\
\vspace{3cm}

\end{center}

\begin{abstract}
The mass matrices of quarks have a simple structure if expressed
in powers of the small parameter $\sigma=(m_c/m_t)^{1/2}$.
If there is a close relation between quarks and leptons, one
would expect similar structures
for the lepton matrices which involve the
same parameter. To have a specific proposal, the see-saw mechanism
is employed together with the stringent requirement that the
singlet neutrinos carry non-zero generation charges. These charges
then determine the powers of $\sigma$ in the corresponding heavy
neutrino mass matrix. As a result, the neutrino mixing matrix
turns out to be of the bimaximal type. In addition, the mass splitting
of the two lightest neutrinos is found to be tiny, and just of the correct
magnitude necessary for the vacuum solution for solar
neutrinos.
\end{abstract}
\end{titlepage}

\section{Introduction}
Recent experiments on solar and atmospheric neutrinos indicate flavor
mixing of neutrinos, and finite neutrino masses \cite{1}, \cite{2}.
Nevertheless, our knowledge on the mass spectrum and the mixing
of neutrinos is still very limited. Several different scenarios
can be envisaged \cite{3}. In contrast to the neutrino properties, 
the masses of the quarks
and the pattern of their mixings are well known. On a qualitative
level only the validity and shape of the unitarity triangle is
still an open problem. In this situation one may ask whether or not
there exists a close relation between the masses and mixings of leptons
and those of quarks. If an intimate relation could be discovered
it would greatly reduce the number of free parameters and would shed
light on the new physics behind the standard model of particle physics.
Such a relation cannot be a trivial one since the large mixing angle
observed in the study of atmospheric neutrinos \cite{2}
has no counterpart in the quark sector.

The mass matrices of quarks, if expressed in powers of a small
parameter, have a simple structure, suggestive of a new generation
symmetry \cite{4}. A connection between quarks and leptons would
be manifest, if the mass matrices of leptons are governed by
powers of the same small parameter. In this work a very close
relation of this type will be proposed. To this end two general
suggestions of grand unified theories \cite{5} are used: the
existence of 3 two-component neutrino fields which are singlets
with respect to the standard model gauge groups, and the similarity
of the Dirac neutrino and charged lepton mass matrices with the
up-quark and the down-quark mass matrix, respectively \cite{6}, \cite{7}.
(But a full gauge theory ($SO(10), E(6)$) with a necessarily
complicated Higgs structure is not implied here, even though
$SO(10)$ models gave the first hint for a large neutrino
mixing angle \cite{6}, \cite{8}). The remaining and decisive mass
matrix is then the one for the 3 singlet neutrinos. Since these
fields are not protected by a standard model symmetry, their
mass values will be very large.

Because of the Majorana type selfcoupling of the singlet neutrinos, 
their generation
charges play an important role in restricting the structure of
the corresponding mass matrix. After exploring these restrictions,
the straightforward use of the see-saw mechanism finally gives
a rather definite form for the mass matrix of the light neutrinos
\cite{9}. It has interesting properties: i) The mixing matrix
obtained from it is of the bimaximal type and also includes
CP-violating phases. ii) By fixing the scale for the singlet
neutrinos to obtain the correct mass scale for the atmospheric
neutrinos,
the mass splitting of the two lightest neutrinos turns out to
be tiny $(\Delta m^2\approx 10^{-11}-10^{-10}({\rm eV})^2)$,
thus favoring the vacuum oscillation solutions \cite{10} for solar
neutrinos.

\section{Quark mass matrices}

The 3x3 mass matrices of up and down quarks contain 36 parameters,
but only 10 parameters are physical: 6 eigenvalues, 3 mixing angles
and 1 CP-violating phase.  Therefore, for different choices of independent
parameters the mass matrices can look very different. From the point
of view of assigning generation (familiy) charges to the quark fields
it is preferable to use a basis in which each independent matrix
element is represented by a different power of a small parameter. A
natural choice for this parameter is $\sigma=(m_c/m_t)^{1/2}=
0.058\pm0.004$, because the top quark plays a dominant role among
the quarks. Moreover, within experimental error limits, one finds
for the up-quark mass ratios
\be\label{1}
m_u:m_c:m_t=\sigma^4:\sigma^2:1\ee
and for the Kobayashi-Maskawa matrix elements \cite{11}
\be\label{2}
|V_{us}|=4\sigma,|V_{cb}|=\sigma/\sqrt2\quad,\quad |V_{ub}|
=\sigma^2\quad.\ee
A suitable basis which involves just 5 parameters for the up-quark mass
matrix $m_U$ and 5 parameters for the down-quark mass matrix
$m_D$, is to take $m_U$ real and symmetric with $(m_U)_{11}=0$ and $m_D$
hermitian with $(m_D)_{13}=(m_D)_{31}=(m_D)_{23}=
(m_D)_{32}=0$. The only complex matrix element is then $(m_D)_{12}
=(m_D)_{21}^*$. The observed masses and mixing angles are well 
represented by taking
\be\label{3}
m_U=\left(\begin{array}{ccc}
0&\sigma^3/\sqrt2&\sigma^2\\
\sigma^3/\sqrt2&-\sigma^2/2&\sigma/\sqrt2\\
\sigma^2&\sigma/\sqrt2&1\end{array}\right)m_t,\quad m_D=
\left(\begin{array}{ccc}
0&-i\sqrt2\sigma^2&0\\
i\sqrt2\sigma^2&-\sigma/3&0\\
0&0&1\end{array}\right)m_b\quad\ee
and $\sigma=0.057$.
However, for our purpose, it is convenient to transform to a basis
in which $m_U$ is diagonal and $m_D=V\ m_D^{diagonal}V^\dagger$,
where $V$ denotes the Cabibbo-Kobayashi-Maskawa matrix. 
To leading order in $\sigma$ one finds\be\label{4}
m_U=\left(\begin{array}{ccc}
\sigma^4&0&0\\
0&\sigma^2&0\\
0&0&1\end{array}\right)\quad m_D=\left(\begin{array}{ccc}
O(\sigma^3)&-i\sqrt2\sigma^2&\sigma^2\\
i\sqrt2\sigma^2&-\sigma/3&\sigma/\sqrt2\\
\sigma^2&\sigma/\sqrt2&1\end{array}\right)m_b\quad.\ee
For definiteness ``maximal CP-violation'' has been used \cite{7,12}.
It is defined to maximize the area of the unitarity triangle with
regard to phase changes of the off-diagonal 
elements \footnote{$m_D$ in (\ref{4}) can also be written in an equivalent
form in which the off-diagonal matrix elements form an 
antisymmetric hermitian matrix \cite{12}.}. 
It leads to a
right-handed unitarity triangle with angles $\alpha\approx70^o,\
\beta\approx 20^o$ and $\gamma\approx90^o$ \cite{11,12}.
Irrespective of this choice the expressions (\ref{3}) and (\ref{4}) 
demonstrate
that masses and mixings are governed by the same small
parameter in a simple fashion.

\section{The Dirac neutrino and charged lepton mass matrices}
The 6x6 neutrino mass matrix (with zero entries for
the light-light sector) has a block structure which contains
a 3x3 Dirac neutrino mass matrix $m_\nu^{Dirac}$ and the
3x3 mass matrix $M$ for the 3 singlet fields $\hat\nu_e,\hat\nu_\mu,
\hat\nu_\tau$\footnote{All fields are taken to be left-handed
two-component fields.}. Since there is no compelling theory
which fixes these matrices and the charged lepton matrix
$m_E$, one is forced to make assumptions. To have a very close
connection between quark and leptons, and to be in line with
the expectations from grand unified theories mentioned in section 1,
I will assume that the mass matrices of quarks and leptons of the
same isospin can be diagonalized simultaneously. For hermitian
forms this implies
\be\label{5}
[m_\nu^{Dirac},\ m_U]=0\qquad[m_E,\ m_D]=0\quad.\ee
More specifically, I will take
\be\label{6}
m_\nu^{Dirac}=m_U\ee
and comment on deviations from (\ref{6}) in
section 5. These equations are assumed to hold at the scale of the
heavy neutrinos.

Eq. (\ref{5}) fixes the mass matrix for the charged leptons
since $m_D$ and the eigenvalues of $m_E$ are known. To illustrate its form
in terms of the parameter $\sigma$, one gets from $m_e:-m_\mu:m_\tau
=\frac{3}{2}\sigma^3:-\sigma:1$ and $m_E=V\ m_E^{Diagonal}V^\dagger$
to leading order in $\sigma$
\be\label{7}
m_E=\left(\begin{array}{ccc}
O(\sigma^3)&-i3\sqrt2\sigma^2&\sigma^2\\
i3\sqrt2\sigma^2&-\sigma&\sigma/\sqrt2\\
i\sigma^2&\sigma/\sqrt2&1\end{array}\right)m_\tau\quad.\ee
Notably, in the basis in which the up-quark mass matrix is
diagonal, the matrix for charged leptons contains complex
elements like $m_D$.

\section{The mass matrix for the singlet neutrinos}
It remains to discuss the mass matrix $M$ for the
singlet neutrino fields $\hat\nu_e, \hat\nu_\mu,\hat\nu_\tau$.
It determines the mass matrix for the light neutrinos $m_\nu$
through the see-saw mechanism
\be\label{8}
m_\nu=-m_U\cdot M^{-1}\cdot m^T_U\quad.\ee
This equation, in which (\ref{6}) has already been used, is supposed
to hold at the scale of the heavy neutrinos. Several suggestions for $M$
can be found in the literature \cite{13}.
A close connection between charged and neutral fermions will exist if
also $M$, in the basis in which $m_U$ is diagonal, has a simple
structure in terms of powers of the small parameter $\sigma$
which occurs in (\ref{4}), (\ref{6}), and (\ref{7})\ \cite{9}.
These powers are restricted if the particle fields carry charges
of a generation (family) symmetry \cite{14}.

Because of the self-coupling of the singlet fields, strong
restrictions for the structure of $M$ occur if the three $U(1)$
generation charges are all different from zero (and $M$ does
not vanish in the limit $\sigma\to 0$)
\cite{9}. By accepting this condition, two of the three fields
must carry opposite $U(1)$ charges. They then form, for
$\sigma\to 0$, a heavy singlet
neutrino of the Dirac type. Giving opposite
charges to $\hat\nu_e$ and $\hat\nu_\tau$ and allowing
integer powers of $\sigma^2$ only, the matrix $M$ has the
structure\footnote{As long as we dismiss matrices $M$ which lead
to a vanishing determinant when neglecting
higher powers of $\sigma$ than $\sigma^2$, the form (\ref{9})
is unique apart from its mirror form corresponding to
the charges $-3/2, -1/2, 3/2$.}
\be\label{9}
M=\left(\begin{array}{ccc}
\sim\sigma^6&\sim\sigma^2&1\\
\sim\sigma^2&\sim\sigma^2&\sim\sigma^4\\
1&\sim\sigma^4&\sim\sigma^6\end{array}\right)M_0.\ee
The $U(1)$ charges for $\hat\nu_e,\hat\nu_\mu,\hat\nu_\tau$
are $-3/2, 1/2, 3/2$, respectively; they determine the powers of 
$\sigma^2$ in (\ref{9}). As in the case of the quark mass
matrices, the proportionality factors multiplying the powers 
of $\sigma$ should
be of order 1. In particular, if the correlation with $m_U$ is
strict, the factor of $\sigma^2$ in the first row and
first column ($p$ in eq. (\ref{10}) should be equal or very
close to one. Because of the smallness of $\sigma^4$ and $\sigma^6$,
$M$ can be approximated by the simpler form
\be\label{10}
M=\left(\begin{array}{ccc}
0&p\sigma^2&1\\
p\sigma^2&r\sigma^2&0\\
1&0&0\end{array}\right)M_0.\ee
One can check that the approximation (\ref{10}) is also
applicable when calculating the mass matrix $m_\nu$ for the
light neutrinos according to (\ref{8}), even though the inverse of the
matrix $M$ enters here. Moreover, a simple consideration of the
original $6\times 6$ neutrino mass matrix (with zero entries
in the light-light sector) shows that the coefficients $p$ and $r$
can be taken to be real.

\section{The mass spectrum and the mixings of the light neutrinos}

The consequences of the simple model considered here are now easily
worked out. From Eqs. (\ref{4}, \ref{8}, \ref{10}) one finds
for the mass matrix of the light neutrinos
\be\label{11}
m_\nu=-\left(\begin{array}{ccc}
0&0&r\sigma^2\\
0&1&-p\\
r\sigma^2&-p&p^2\end{array}\right)
\frac{\sigma^2m_t(M_0)^2}{rM_0}~.\ee
Apart from the mass scale $M_0$ we are left with only two important
parameters $(p,r)$. The remarkable point about this matrix
is that it leads to a near degeneracy of the two lightest neutrinos,
and for $p\approx 1$, to a neutrino mixing matrix which is of the
bimaximal type as defined in \cite{15}. To make contact
with the atmospheric neutrino data \cite{2} the heavy mass parameter
$M_0$ can be adjusted to give the heaviest of the light neutrinos
$(\nu_3)$ the mass $m_3\approx 5.5\times10^{-2}$ eV. For
$p=r=1$ one then finds $M_0= 10^{12}$ GeV,
$m^2_3-m^2_1=3\times 10^{-3}({\rm eV})^2$
and $m^2_2-m^2_1=1\times 10^{-11}$ (eV)$^2$.
For $r=2$ one gets $M_0=5\times 10^{11}$ GeV and $m^2_2-m^2_1
=7\times 10^{-11}({\rm eV})^2$. Thus, the neutrino mass matrix
obtained by invoking an intimate relation between charged and
neutral fermions favors large mixing angles for the atmospheric
and for the solar neutrinos, and the vacuum oscillation solution
\cite{10} for the latter. But it seems not compatible with
indications for $\bar\nu_\mu\to\bar\nu_e$ oscillations reported by
the LSND collaboration \cite{16}.

For a more detailed treatment one has, of course, to take
the charged lepton mass matrix $m_E$, which is nondiagonal
in our basis, into account. In addition, the scale dependence
of $m_\nu$ according to the renormalization group has to be
considered.

Transforming therefore to a basis in which $m_E$ is diagonal, the
corresponding neutrino mass matrix at the scale $M_0$ is then
\be\label{12}
\tilde m_\nu(M_0)=V^T(M_0)m_\nu(M_0)V(M_0)~.\ee
For $m_\nu(M_0)$ we have to replace (\ref{11}) by
\be\label{13}
m_\nu(M_0)=-\left(\begin{array}{ccc}
0&0&r\frac{m_u(M_0)}{m_c(M_0)}\\
0&1&-p\\
r\frac{m_u(M_o)}{m_c(M_0)}&-p&p^2\end{array}\right)
\frac{m_c(M_0)m_t(M_0)}{rM_0}~~.\ee
Now, the renormalization group equation  for $\tilde m_\nu
(\mu)$ \cite{17}
can be used to calculate $\tilde m_\nu$ at the weak scale $m_Z$.
After this calculation the neutrino mixing matrix $U$ is obtained
by diagonalizing the hermitian matrix $\tilde m_\nu\cdot \tilde m^*_\nu$:
\be\label{14}
\tilde m_\nu(m_Z)\cdot\tilde m_\nu^*(m_Z)=UDD^*U^\dagger\ .\ee
The diagonal matrix $D$
\be\label{15}
D=U^\dagger\tilde m_\nu(m_Z)U^*\ .\ee
gives the neutrino mass eigenvalues. These are complex because
of the complex Cabibbo-Kobayashi-Maskawa matrix elements. Finally,
$U$ can be redefined by a diagonal phase matrix such that (\ref{15})
contains now only positive definite eigenvalues. The new matrix
$U$ then expresses the light neutrino states $\nu_e, \nu_\mu,
\nu_\tau$ in terms of the neutrino mass eigenstates $\nu_1,\nu_2,\nu_3$
according to
\be\label{16}
\left(\begin{array}{c} \nu_e\\ \nu_\mu\\\nu_\tau\end{array}\right)
=U\left(\begin{array}{c}\nu_1\\ \nu_2\\
\nu_3\end{array}\right)\quad.\ee
From the knowledge of $U$, the surviving and transition
probabilities of the oscillating neutrino flavors can easily
be calculated.

It turns out that the mass spectrum and mixings obtained from
(\ref{14}),(\ref{15}) is not strikingly different from the ones found
directly from the simple mass matrix, eq. (\ref{11}). The small
mixing angles occurring in $V$ do not strongly influence $\tilde m_\nu$.
The mixing matrix $U$ has again -- for $p\approx 1$ -- a near bimaximal
form, but contains now also CP-violating contributions \cite{9}.
Also the scale change from $M_0\approx 10^{12}$ GeV down to the
weak scale does not lead to a qualitatively different picture.
In particular, the near degeneracy of the two lightest neutrino
masses remains stable \cite{9} in contrast to cases discussed
in ref. \cite{18}.

To illustrate the results I give here the actual numbers taking 
$p=1$ and $r=2$. (They differ somewhat from the ones obtained in
\cite{9} since I use now in (\ref{12}) the 
Cabibbo-Kobayashi-Maskawa Matrix $V$ 
as a concequence of the condition (\ref{5})). 
With the input $m_3 = 0.055~ eV$
one gets $M_0=3\times 10^{11}~ GeV $, $m_3^2-m_1^2=3\times 10^{-3}
(eV)^2~ $and $m_2^2-m_1^2= 9\times
10^{-11} (eV)^2$. The survival and transition probabilities are
(for notations see \cite{9})
\begin{eqnarray}\label{17}
P(\nu_e \rightarrow \nu_e)&=&1-0.95~ S_{21}-0.05 ~S_{31}-
0.04 ~S_{32}  \nonumber\\
P(\nu_{\mu} \rightarrow \nu_{\mu})&=&1-0.32 ~S_{21}-
0.49 ~S_{31}-0.49 ~S_{32}  \nonumber\\
P(\nu_{\tau} \rightarrow \nu_{\tau})&=&1-0.21~S_{21}-
0.49~S_{31}-0.50~S_{32}  \nonumber\\ 
P(\nu_e \rightarrow \nu_{\mu})&=
&0.53 ~S_{21}+0.02 ~S_{31}+0.02 ~S_{32}
\nonumber\\&+& 0.07 ~T_{21}
- 0.07 ~T_{31}+0.07 ~T_{32}  \nonumber\\
P(\nu_e \rightarrow \nu_{\tau})&=
&0.42 ~S_{21}+0.02 ~S_{31}+0.03 ~S_{32}
 \nonumber\\&-& 0.07 ~T_{21}+
0.07 ~T_{31}-0.07 ~T_{32}  \nonumber\\
P(\nu_{\mu} \rightarrow \nu_{\tau})&=
&-0.21 ~S_{21}+0.47 ~S_{31}+0.47 ~S_{32}
\nonumber\\ &+& 0.07 ~T_{21}-
 0.07 ~T_{31}+0.07 ~T_{32}~~.
\end{eqnarray}

Our mass matrix for the light neutrinos is based on
the eqs. (\ref{5}), (\ref{6}) and (\ref{13}). If the eigenvalues 
of $m^{Dirac}_\nu$
differ from those of the up-quark mass matrix, but (\ref{5}) still holds,
the corresponding modification can be absorbed into the parameters
$r$ and $M_0$, which have anyhow to be adjusted to the data.
A large effective value for $r$ would give a significantly larger 
value for
$m_2^2-m_1^2$ . In this case the model predicts 
an energy independent deficit of solar neutrinos, which is,
however, barely compatible with the results of the Homestake
collaboration \cite{1}.

The bimaximal mixing depends on the parameter $p$ and gets spoiled
if $p$ is sizeably different from 1. The mixing angle for
atmospheric neutrinos is sensitive to $p$. Deviations from
$p=1$ by more than 25 \% would no more be compatible with
the mixing angle extracted from present experiments.

\section{Conclusion}
A close relationship between leptons and quarks has been proposed.
All mass matrices are expressed in terms of powers of the same
small parameter $\sigma=(m_c/m_t)^{1/2}$. To obtain the neutrino
mass matrix, the see-saw mechanism is employed. Rather detailed
predictions on the spectrum and the mixing angles follow from
an Ansatz for the mass matrix of the singlet neutrino fields,
which is based on the assumption of non-zero generation (family)
charges of these particles. The mixing matrix is of the bimaximal
type, and the mass difference of the two lightest neutrinos has the 
right magnitude required for the 
vacuum oscillation solution for solar neutrinos. The spectrum and 
the mixing pattern is stable with respect to loop corrections.

\section*{Acknowledgement}
The author likes to thank Susan and Gabor Domokos for
the invitation to this workshop and for their kind hospitality.
He also likes to thank his colleague Christof Wetterich for
very helpful comments.


\begin{thebibliography}{12}
\bibitem{1} For the results of the collaborations
Gallex, Homestake, Sage, Kamiokande
and Superkamiokande see Bahcall et al., hep-ph 9807216.
\bibitem{2} Y. Fukuda et al., hep-ex/9807003.
\bibitem{3} A. Yu Smirnov, hep-ph/9901208;\\
W. M. Alberico and S. M. Bilenky, hep-ph/9905254;\\
P. Fisher, B. Kayser and K. S. McFarland, NSF-PT-99-1;\\
G. Altarelli and F. Feruglio, hep-ph/9905536.
\bibitem{4} C. D. Froggatt and H. B. Nielsen, Nucl. Phys.
{\bf B147} (1979) 277;\\
S. Dimopoulos, Phys. Lett. {\bf B129} (1983) 417;\\
J. Bijnens and C. Wetterich, Nucl. Phys. {\bf B283} (1987) 237;\\
Phys. Lett. {\bf B199} (1987) 525;\\
L. E. Ibanez and G. G. Ross, Phys. Lett. {\bf B332} (1994) 100.
\bibitem{5} see G. G. Ross, ``Grand Unified Theories'', Benjamin 1985.
\bibitem{6} J. A. Harvey, P. Ramond, and D. B. Reiss, Nucl. Phys.
{\bf B199} (1982) 223.
\bibitem{7} B. Stech, in: Flavor Mixing in Weak Interactions,
Ed. L. L. Chau, Plenum (1984) p. 735.
\bibitem{8} J. Bijnens and C. Wetterich, Nucl. Phys. {\bf B292} (1987)
443.
\bibitem{9} B. Stech, hep-ph/9905440
\bibitem{10} J. H. Bahcall, P. I. Krastev, and A. Yu. Smirnov,
Phys. Rev. {\bf D58} (1998) 096016;\\
V. Barger and K. Whisnant, hep-ph/9903262.
\bibitem{11} B. Stech, Phys. Lett. {\bf B403} (1997) 114.
\bibitem{12} B. Stech, Phys. Lett. {\bf B130} (1983) 189
\bibitem{13} e.g. M. Jezabek and Y. Sumino, Phys. Lett. {\bf B440}
(1998) 327, hep-ph/9807310, hep-ph/9904382;\\
T. Blazek, S. Raby, and K. Tobe, hep-ph/9903340;\\
C. H. Albright and S. M. Barr, Phys. Lett. {\bf B452} (1999) 287;\\
R. Barbieri, P. Creminelli, and A. Romanino, hep-ph/9903460.
\bibitem{14} N. Irges, S. Lavignac, and P. Ramond, Phys. Rev. {\bf D58}:  
(1998) 035003;\\
J. K. Ellwood, N. Irges, and P. Ramond, Phys. Rev. Lett. {\bf 81}:
(1998) 5064 ;\\
G. Altarelli and F. Feruglio, Phys. Lett. {\bf B451} (1999) 388;\\
S. Lola, hep-ph/9903203.
\bibitem{15} V. Barger, S. Pakvasa, T. J. Weiler, and K. Whisnant,
Phys. Lett. {\bf B437} (1998) 107.
\bibitem{16}
C. Athanassopoulos et al., LSND Coll., Phys. Rev. {\bf C58} (1998) 2511.
\bibitem{17} C. Wetterich, Nucl. Phys. {\bf B187} (1981) 343;\\
K. Babu, C. N. Leung, and J. Pantalone, Phys. Lett. {\bf B319}
(1993) 191.
\bibitem{18}
J. Ellis and S. Lola, hep-ph/9904279.


\end{thebibliography}
\end{document}